\documentclass[aps,twocolumn,showpacs,preprintnumbers,amsmath,amssymb,nofootinbib,superscriptaddress,showkeys]{revtex4-1}

\usepackage{hyperref}
\usepackage{float}
\usepackage{epsfig}
\usepackage{graphicx}
\usepackage{mathptmx}      % use Times fonts if available on your TeX system
\usepackage{latexsym}
\usepackage{natbib}
\begin{document}

\title{Partial Wave Analysis of Nucleon-Nucleon Scattering below pion
  production threshold}

\author{R. Navarro P\'erez}\email{rnavarrop@ugr.es}
\affiliation{Departamento de F\'{\i}sica At\'omica, Molecular y
  Nuclear \\ and Instituto Carlos I de F{\'\i}sica Te\'orica y
  Computacional \\ Universidad de Granada, E-18071 Granada, Spain.}
\author{J.E. Amaro}\email{amaro@ugr.es} \affiliation{Departamento de
  F\'{\i}sica At\'omica, Molecular y Nuclear \\ and Instituto Carlos I
  de F{\'\i}sica Te\'orica y Computacional \\ Universidad de Granada,
  E-18071 Granada, Spain.}  \author{E. Ruiz
  Arriola}\email{earriola@ugr.es} \affiliation{Departamento de
  F\'{\i}sica At\'omica, Molecular y Nuclear \\ and Instituto Carlos I
  de F{\'\i}sica Te\'orica y Computacional \\ Universidad de Granada,
  E-18071 Granada, Spain.}

\date{\today}

\begin{abstract} 
\rule{0ex}{3ex} We undertake a simultaneous partial wave analysis to
proton-proton and neutron-proton scattering data below pion production
threshold up to LAB energies of $350 {\rm MeV}$.  We represent the
interaction as a sum of delta-shells in configuration space below 
$3 {\rm fm}$ and a charge dependent one pion exchange potential above
$3 {\rm fm}$ together with electromagnetic effects.  We obtain
$\chi^2= 2813|_{pp} + 3985|_{np} $ with a total of $N=2747|_{pp} +
3691|_{np}$ data obtained till 2013 and a total number of $46$ fitting
parameters yielding $\chi^2/{\rm d.o.f} = 1.06$. Special attention is
paid to estimate the errors of the phenomenological interaction as
well as the derived effects on the phase-shifts and scattering
amplitudes.
\end{abstract}


\maketitle

\section{Introduction}
%\label{sec:intro}

The NN interaction plays a central role in Nuclear Physics (see
e.g.~\cite{Machleidt:1989tm,Machleidt:2011zz} and references
therein). 
%While there have been several attempts to construct it
%realistically from lattice QCD, the most accurate procedure to date
The standard procedure to constrain the interaction uses a partial
wave analysis (PWA) of the proton-proton (pp) and neutron-proton (np)
scattering data below pion production
threshold~\cite{glöckle1983quantum} although there are accurate
descriptions up to $3{\rm GeV}$ for pp and $1.3{\rm GeV}$ for np~\cite{Arndt:2007qn}. The Nijmegen PWA uses
a large body of NN scattering data giving $\chi^2 /{\rm d.o.f}
\lesssim 1 $ after discarding about $20\%$ of $3\sigma$ inconsistent
data~\cite{Stoks:1993tb} (see however~\cite{Arndt:2007qn} where
$\chi^2 /{\rm d.o.f} = 1.4 $ without the $3\sigma$
criterium). This fit incorporates charge dependence (CD) for the One
Pion Exchange (OPE) potential as well as electromagnetic, vacuum
polarization and relativistic effects, the latter being key
ingredients to this accurate success. The analysis was more
conveniently carried out using an energy dependent potential for the
short range part. Later on energy independent {\it high quality}
potentials were designed with almost identical $\chi^2/{\rm d.o.f}
\sim 1$ for the gradually increasing
database~\cite{Stoks:1994wp,Wiringa:1994wb,Machleidt:2000ge,Gross:2008ps}.
While any of these potentials provides individually satisfactory fits
to the available experimental data, an error analysis would add a
means of estimating quantitatively the impact of NN-scattering
uncertainties in Nuclear Structure calculations. In the present work
we provide a new high-quality potential implementing an analysis of
its parameter uncertainties using the standard method of inverting the
covariance matrix~\cite{taylor-book}.

The work is organized as follows. In Section~\ref{sec:2} we briefly
review the main aspects of the formalism. After that, in
Section~\ref{sec:3} we present our numerical results and fits as well
as our predictions for deuteron properties and scattering amplitudes.
Finally, in Section~\ref{sec:4} we summarize our results and come to
the conclusions.

\begin{table*}[htb]
	\centering
      \caption{Fitting delta-shell parameters $(\lambda_n)^{JS}_{l,l'}
        $ (in ${\rm fm}^{-1}$) with their errors for all states in the
        $JS$ channel. We take $N=5$ equidistant points with $\Delta r
        = 0.6$fm. $-$ indicates that the corresponding fitting 
        $(\lambda_n)^{JS}_{l,l'} =0$. In the first line we provide the
        central component of the delta shells corresponding to the EM
        effects below $r_c = 3 {\rm fm}$. These parameters remain
        fixed within the fitting process.}
      \label{tab:Fits}
	\begin{tabular*}{\textwidth}{@{\extracolsep{\fill}}c c c c c c c}
            \hline
            \hline\noalign{\smallskip}
		Wave  & $\lambda_1$ & $\lambda_2$ & $\lambda_3$ & $\lambda_4$ & $\lambda_5$  \\

	   & ($r_1=0.6$fm) & ($r_2=1.2$fm) & ($r_3=1.8$fm) & ($r_4=2.4$fm) & ($r_5=3.0$fm)  \\
	
%		  & & fm & fm & fm$^{-1}$ & fm$^{-1}$ & fm$^{-1}$ &  \\
            \hline\noalign{\smallskip}
$V_C[{\rm pp}]_{\rm EM}$  & 0.02091441 & 0.01816750 & 0.00952244
& 0.01052224
& 0.00263887 \\ 
\hline 
             $^1S_0{\rm np}$& 1.28(7)  & -0.78(2)  & -0.16(1)  &      $-$   & -0.025(1)  \\
             $^1S_0[\rm pp]$& 1.31(2)  & -0.723(4) & -0.187(2) &      $-$   & -0.0214(3) \\
             $^3P_0$        &     $-$  &  1.00(2)  & -0.339(7) & -0.054(3)  & -0.025(1)  \\
             $^1P_1$        &     $-$  &  1.19(2)  &      $-$  &  0.076(2)  &      $-$   \\
             $^3P_1$        &     $-$  &  1.361(5) &      $-$  &  0.0579(5) &      $-$   \\
             $^3S_1$        & 1.58(6)  & -0.44(1)  &      $-$  & -0.073(1)  &      $-$   \\
             $\varepsilon_1$&     $-$  & -1.65(1)  & -0.34(2)  & -0.233(8)  & -0.020(3)  \\
             $^3D_1$        &     $-$  &      $-$  &  0.35(1)  &  0.104(9)  &  0.014(3)  \\
             $^1D_2$        &     $-$  & -0.23(1)  & -0.199(3) &      $-$   & -0.0195(2) \\
             $^3D_2$        &     $-$  & -1.06(4)  & -0.14(2)  & -0.243(6)  & -0.019(2)  \\
             $^3P_2$        &     $-$  & -0.483(1) &      $-$  & -0.0280(6) & -0.0041(3) \\
             $\varepsilon_2$&     $-$  &  0.28(2)  &  0.200(4) &  0.046(2)  &  0.0138(5) \\
             $^3F_2$        &     $-$  &  3.52(6)  & -0.232(4) &      $-$   & -0.0139(6) \\
             $^1F_3$        &     $-$  &      $-$  &  0.13(2)  &  0.091(8)  &      $-$   \\
             $^3D_3$        &     $-$  &  0.52(2)  &      $-$  &      $-$   &      $-$   \\
            \noalign{\smallskip}\hline
            \hline
	\end{tabular*}
\end{table*}

\begin{table*}[htb]
	\centering
	\caption{Deuteron static properties compared with empirical values and high-quality potentials calculations}
	\label{tab:DeuteronP}
	\begin{tabular*}{\textwidth}{@{\extracolsep{\fill}}l l l l l l l l }
      \hline
      \hline
            & Delta Shell & Empirical\cite{VanDerLeun1982261,Borbély198517,Rodning:1990zz,Klarsfeld1986373,Bishop:1979zz,deSwart:1995ui} & Nijm I~\cite{Stoks:1994wp}   & Nijm II~\cite{Stoks:1994wp}  & Reid93~\cite{Stoks:1994wp}   & AV18~\cite{Wiringa:1994wb} & CD-Bonn~\cite{Machleidt:2000ge}  \\
% & Spect~\cite{Gross:2008ps} 0.0256(4) 0.8777(15) missing others
      \hline
		$E_d$(MeV)              & Input       & 2.224575(9)    & Input    & Input    & Input    & Input  & Input           \\
		$\eta$                  & 0.02493(8)  & 0.0256(5)      & 0.02534  & 0.02521  & 0.02514  & 0.0250 & 0.0256            \\
		$A_S ({\rm fm}^{1/2})$  & 0.8829(4)   & 0.8781(44)     & 0.8841   & 0.8845   & 0.8853   & 0.8850  & 0.8846    \\
		$r_m ({\rm fm})$        & 1.9645(9)   & 1.953(3)       & 1.9666   & 1.9675   & 1.9686   & 1.967 &  1.966           \\
		$Q_D ({\rm fm}^{2}) $   & 0.2679(9)   & 0.2859(3)      & 0.2719   & 0.2707   & 0.2703   & 0.270  & 0.270   \\
		$P_D$                   & 5.62(5)     & 5.67(4)        & 5.664    & 5.635    & 5.699    & 5.76  & 4.85              \\
		$\langle r^{-1} \rangle ({\rm fm}^{-1})$& 0.4540(5)    &                &          & 0.4502   & 0.4515   &  & \\
      \hline \hline
	\end{tabular*}
\end{table*}
\begin{figure*}[hpt]
\begin{center}
\epsfig{figure=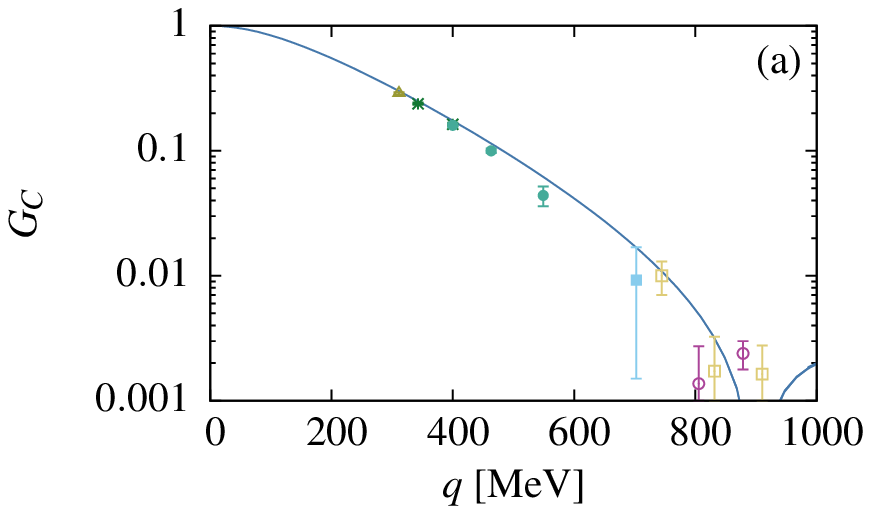,height=4cm,width=5cm}
\epsfig{figure=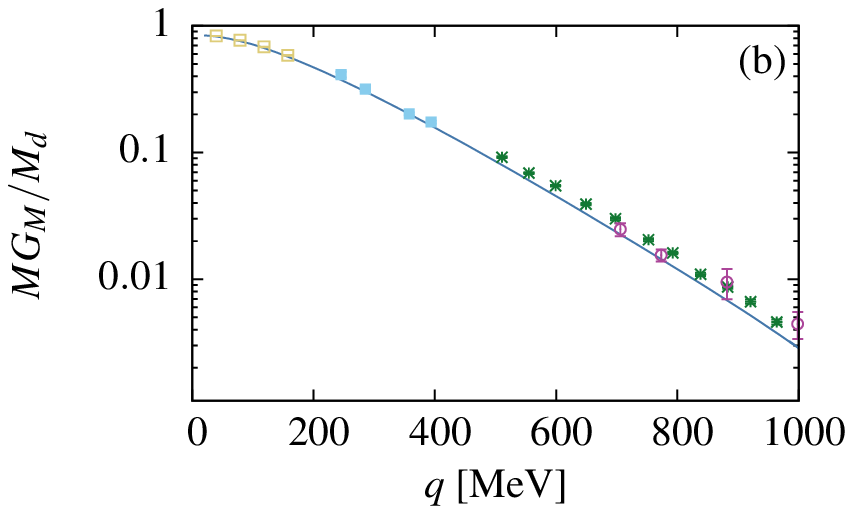,height=4cm,width=5cm}
\epsfig{figure=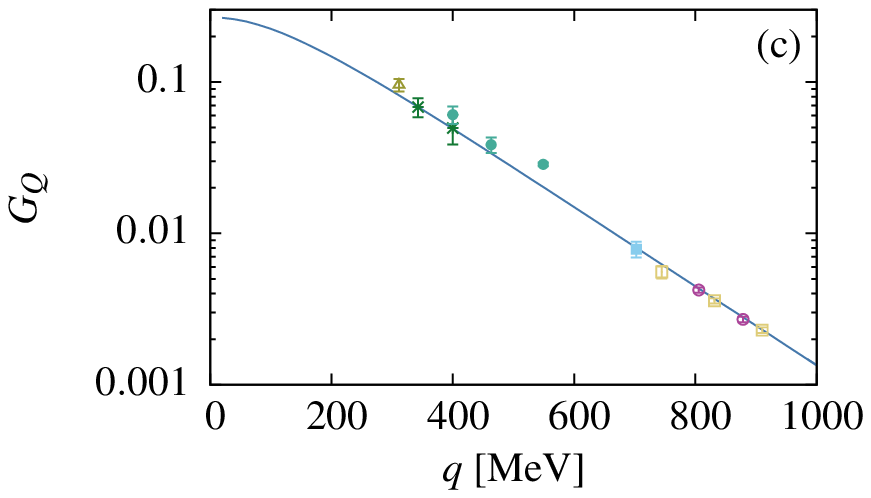,height=4cm,width=5cm}
\end{center}
\caption{Deuteron Form Factors with theoretical error bands obtained
  by propagating the uncertainties of the np+pp plus deuteron binding
  fit (see main text).  Note that the theoretical error is so tiny
  that the width of the bands cannot be seen at the scale of the
  figure.}
\label{FigFormFactors}
\end{figure*}

\section{Formalism}
\label{sec:2}
The complete on-shell NN scattering amplitude contains five
independent complex quantities, which we choose for definiteness as
the Wolfenstein parameters~\cite{glöckle1983quantum}
\begin{eqnarray}
   M(\mathbf{k}_f,\mathbf{k}_i) &=& a + m (\mathbf{\sigma}_1,\mathbf{n})(\mathbf{\sigma}_2,\mathbf{n}) 
                  + (g-h)(\mathbf{\sigma}_1,\mathbf{m})(\mathbf{\sigma}_2,\mathbf{m}) \nonumber \\
                  & &+ (g+h)(\mathbf{\sigma}_1,\mathbf{l})(\mathbf{\sigma}_2,\mathbf{l})  + c(\mathbf{\sigma}_1+\mathbf{\sigma}_2,\mathbf{n}) \, ,  
\end{eqnarray}
where $a,m,g,h,c$ depend on energy and angle, $\mathbf{\sigma}_1$ and
$\mathbf{\sigma}_2$ are the single nucleon Pauli matrices,
$\mathbf{l}$, $\mathbf{m}$, $\mathbf{n}$ are three unitary orthogonal
vectors along the directions of $\mathbf{k}_f+\mathbf{k}_i$,
$\mathbf{k}_f-\mathbf{k}_i$ and $\mathbf{k}_i \wedge \mathbf{k}_f$ and
$\mathbf{k}_f$, $\mathbf{k}_i$ are the final and initial relative
nucleon momenta respectively.  To determine these parameters and their
uncertainties we find that a convenient representation to sample the
short distance contributions to the NN interaction can be written as a
sum of delta-shells
\begin{eqnarray}
   V(r) &=& \sum_{n=1}^{18} O_n \left[\sum_{i=1}^N V_{i,n} \Delta r_i \delta(r-r_i) \right] \nonumber \\ &+& \Big[ V_{\rm OPE}(r) + V_{\rm em}(r) \Big] \theta(r-r_c),
\label{eq:potential}
\end{eqnarray}
where $O_n$ are the set of operators in the AV18
basis~\cite{Wiringa:1994wb}, $r_i \le r_c$ are a discrete set of
$N$-radii, $\Delta r_i = r_{i+1}-r_i $
and $V_{i,n}$ are unknown coefficients to be determined from
data.  The $r > r_c$ piece contains a CD OPE and electromagnetic (EM)
corrections which is kept fixed throughout. The solution of the
corresponding Schr\"odinger equation in the (coupled) partial waves
$^{2S+1}L_J$ for $r \le r_c$ is straightforward since the potential
reads
\begin{eqnarray}
V^{JS}_{l,l'}(r) = \frac{1}{2\mu_{\alpha\beta}}\sum_{i=1}^N  (\lambda_i)^{JS}_{l,l'} \delta(r-r_i)  \qquad r \le r_c
\label{eq:ds-pot}
\end{eqnarray}
with $\mu_{\alpha\beta}=M_\alpha M_\beta/(M_\alpha+M_\beta)$ the
reduced mass with $\alpha,\beta=n,p$.  Here,  $(\lambda_i)^{JS}_{l,l'} $ 
are related to the $V_{i,n}$ coefficients by linear transformation at
each discrete radius $r_i$. Thus, for any $r_i < r < r_{i+1} $ we have
free particle solutions and log-derivatives are discontinuous at the
$r_i$-radii so that one generates an accumulated S-matrix at any
sampling radius providing a discrete and purely algebraic version of
Calogero's variable phase equation~\cite{Calogero:1965}.

This form of potential effectively implements a coarse graining of the
interaction, first proposed 40 years ago by
Aviles~\cite{Aviles:1973ee}.  We have found that the
representation~(\ref{eq:ds-pot}) is extremely convenient and
computationally cheap for our PWA. The low energy expansion of the
discrete variable phase equations was used already in
Ref.~\cite{PavonValderrama:2005ku} to determine threshold parameters
in all partial waves. The relation to the well-known Nyquist theorem
of sampling a signal with a given bandwidth has been discussed in
Ref.~\cite{Entem:2007jg}. Some of the advantages of directly using
this simple potential for Nuclear Structure calculations have also
been analyzed~\cite{NavarroPerez:2011fm}.

The fact that we are coarse graining the interaction enables to encode
efficiently all effects operating below the finest resolution $\Delta
r$ which we identify with the shortest de Broglie wavelength
corresponding to the pion production threshold, $ \lambda_{\rm min}
\sim 1/\sqrt{m_\pi M_N} \sim 0.55 {\rm fm}$, so that a maximal number
of delta-shells $N = r_c / \Delta r \sim 5$ (for $r_c = 3 {\rm fm}$) should
be needed. In practice, we expect the number of sampling radii to
decrease with angular momentum as the centrifugal barrier make
irrelevant those radii $r_i \lesssim (l+1/2)/p $ below the relevant
impact parameter, so that the total number of delta-shells and
hence fitting strengths $V_{i,n}$ will be limited and smaller than
$N=5$.

The previous discretization of the potential is just a way to
numerically solve Schr\"odinger equation for any given potential where
one replaces $V(r) \to \bar V(r) = \sum_i V (r_i) \Delta r_i \delta(r-r_i)$,
but the number of delta-shells may be quite large for {\it fixed}
strengths $V_i \equiv V(r_i)$. For instance, for the $^1S_0$ wave and for the 
AV18~\cite{Wiringa:1994wb} potential one needs N=600
delta-shells to reproduce the phase-shift with sufficient accuracy
(below $10^{-4}$ degrees) but just $N=5$ if one uses $V(r_i)$ as fitting
parameters to the {\it same} phase shift~\cite{NavarroPerez:2011fm}. 
%This is
%unlike fitting continuous representations of the potential where the numerical
%accuracy of the integration step of the differential equation solver
%is determined by the small changes in the wave function.

The EM part of the NN potential gives a contribution to
the scattering amplitude that must be taken into account properly in
order to correctly calculate the different observables. Each term of
the electromagnetic potential in the pp and np channels needs to be
treated differently to obtain the corresponding parts of the total EM
amplitude.  The expresions for the contributions coming from the pp
one photon exchange potential $V_{C1}$, and the corresponding
relativistic correction $V_{C2}$, are well known and can be found
in~\cite{Stoks:1993tb}. To calculate the contribution of the vacuum
polarization term $V_{VP}$ we used the approximation to the amplitude
given in~\cite{Durand:1957zz}.  Finally, Ref.~\cite{Stoks:1990us}
details the treatment of the magnetic moment interaction $V_{MM}$ for
both pp and np channels and the necesary corrections to the nuclear
amplitude coming from the electromagnetic phaseshifts.

\begin{figure*}[hp]
\begin{center}
\epsfig{figure=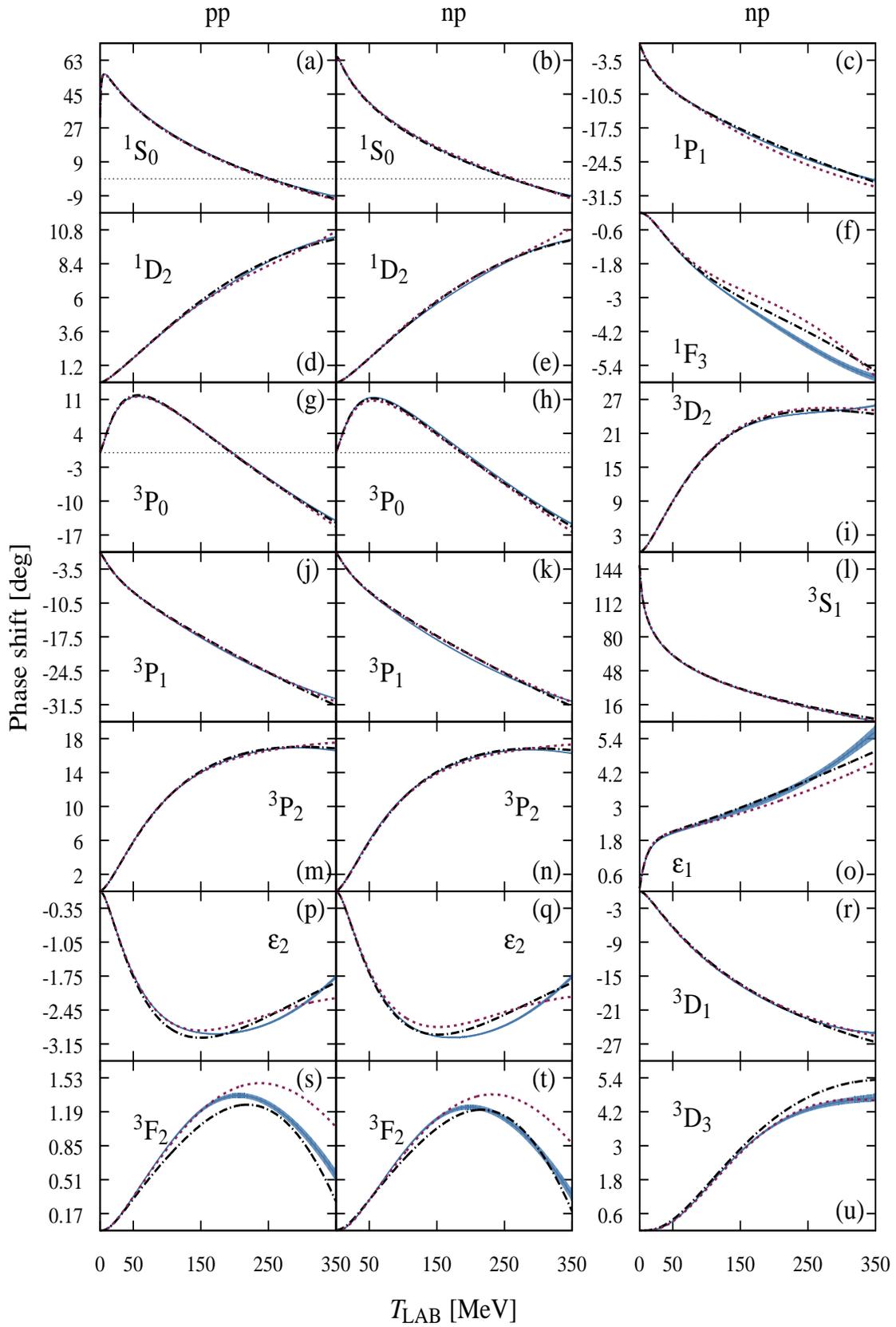,height=22cm,width=14.7cm}
\end{center}
\caption{(Color on-line) np and pp phase shifts and their propagated errors 
(blue band)
  corresponding to independent operator combinations 
  of the fitted potential.  We compare our fit (blue band) with
  the PWA~\cite{Stoks:1993tb} (dotted,magenta) and the AV18
  potential~\cite{Wiringa:1994wb} (dashed-dotted,black) 
  which gave $\chi^2/ {\rm d.o.f} \lesssim 1 $ for data before 1993.}
\label{fig:all-phases}
\end{figure*}

\begin{figure*}[hpt]
\begin{center}
\epsfig{figure=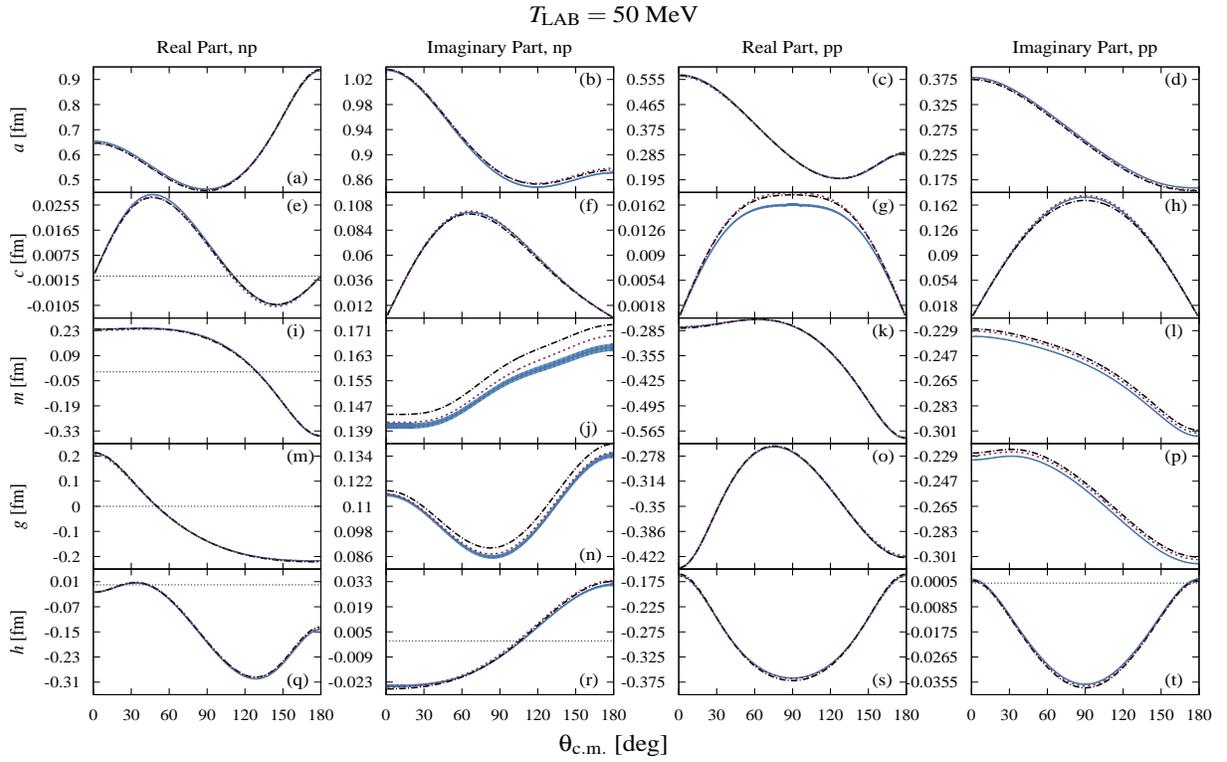,height=10cm,width=16cm}
\end{center}
\caption{Color on-line. np (left) and pp (right) Wolfenstein
  parameters (in fm) as a function of the CM angle (in degrees) and
  for $E_{\rm LAB}=50 {\rm MeV}$. We compare our fit (blue band) with
  the PWA~\cite{Stoks:1993tb} (dotted,magenta) and the AV18
  potential~\cite{Wiringa:1994wb} (dashed-dotted,black) which provided
  a $\chi^2/ {\rm d.o.f} \lesssim 1 $ for data before 1993.}
\label{FigWolfenstein050}
\end{figure*}

\begin{figure*}[hpt]
\begin{center}
\epsfig{figure=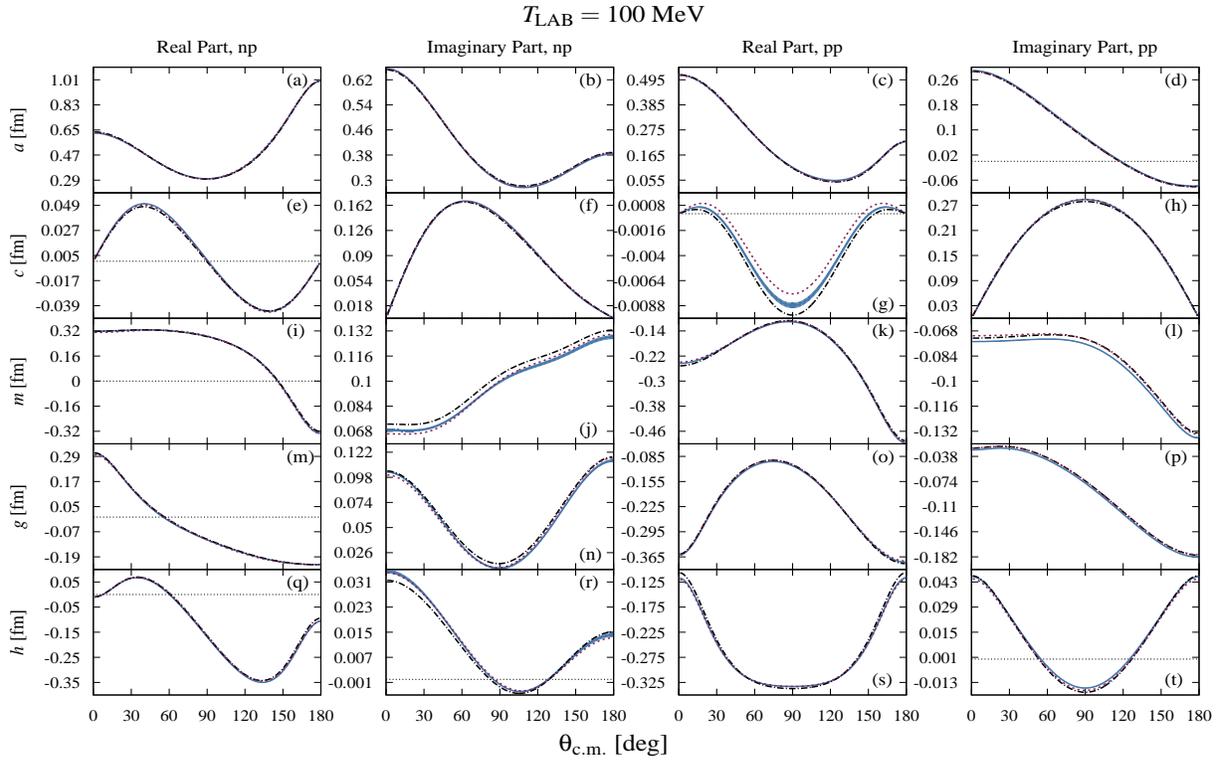,height=10cm,width=16cm}
\end{center}
\caption{Same as in Fig.~\ref{FigWolfenstein050} but for  $E_{\rm LAB}=100 {\rm
    MeV}$.}
\label{FigWolfenstein100}
\end{figure*}

\begin{figure*}[hpt]
\begin{center}
\epsfig{figure=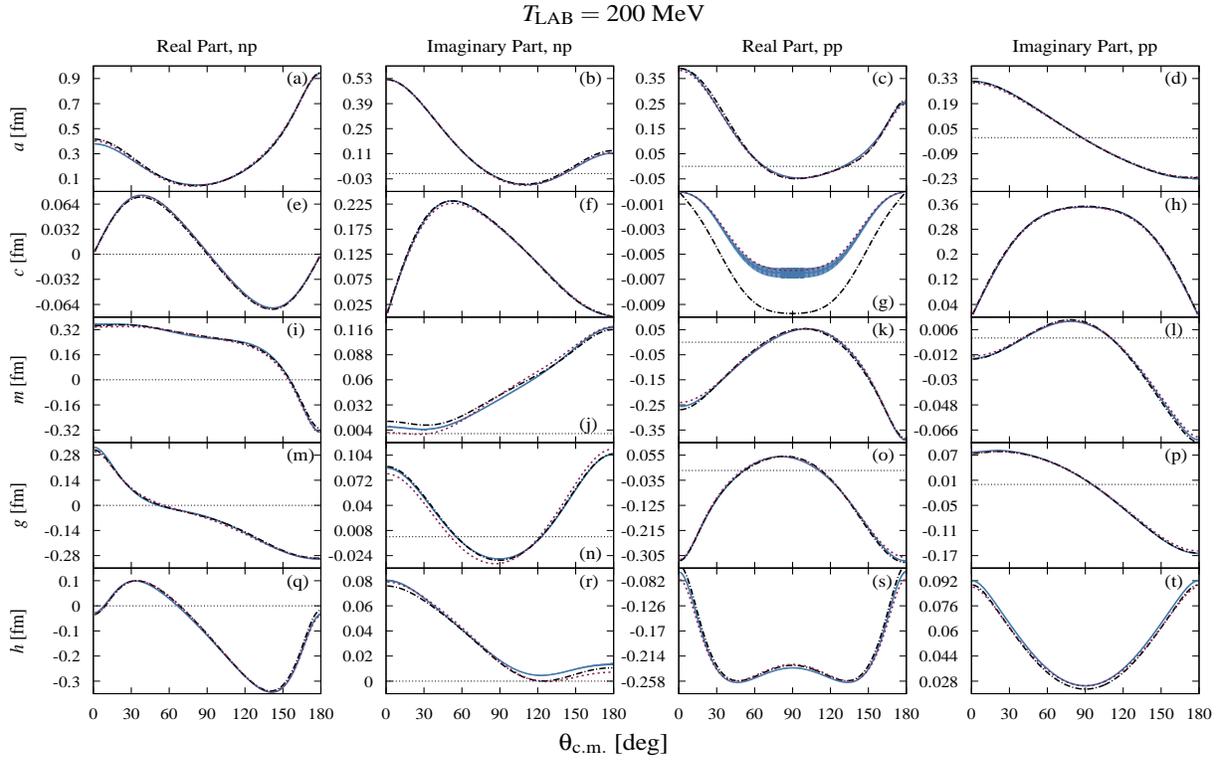,height=10cm,width=16cm}
\end{center}
\caption{Same as in Fig.~\ref{FigWolfenstein050} but for  $E_{\rm LAB}=200 {\rm
    MeV}$.}
\label{FigWolfenstein200}
\end{figure*}

\begin{figure*}[hpt]
\begin{center}
\epsfig{figure=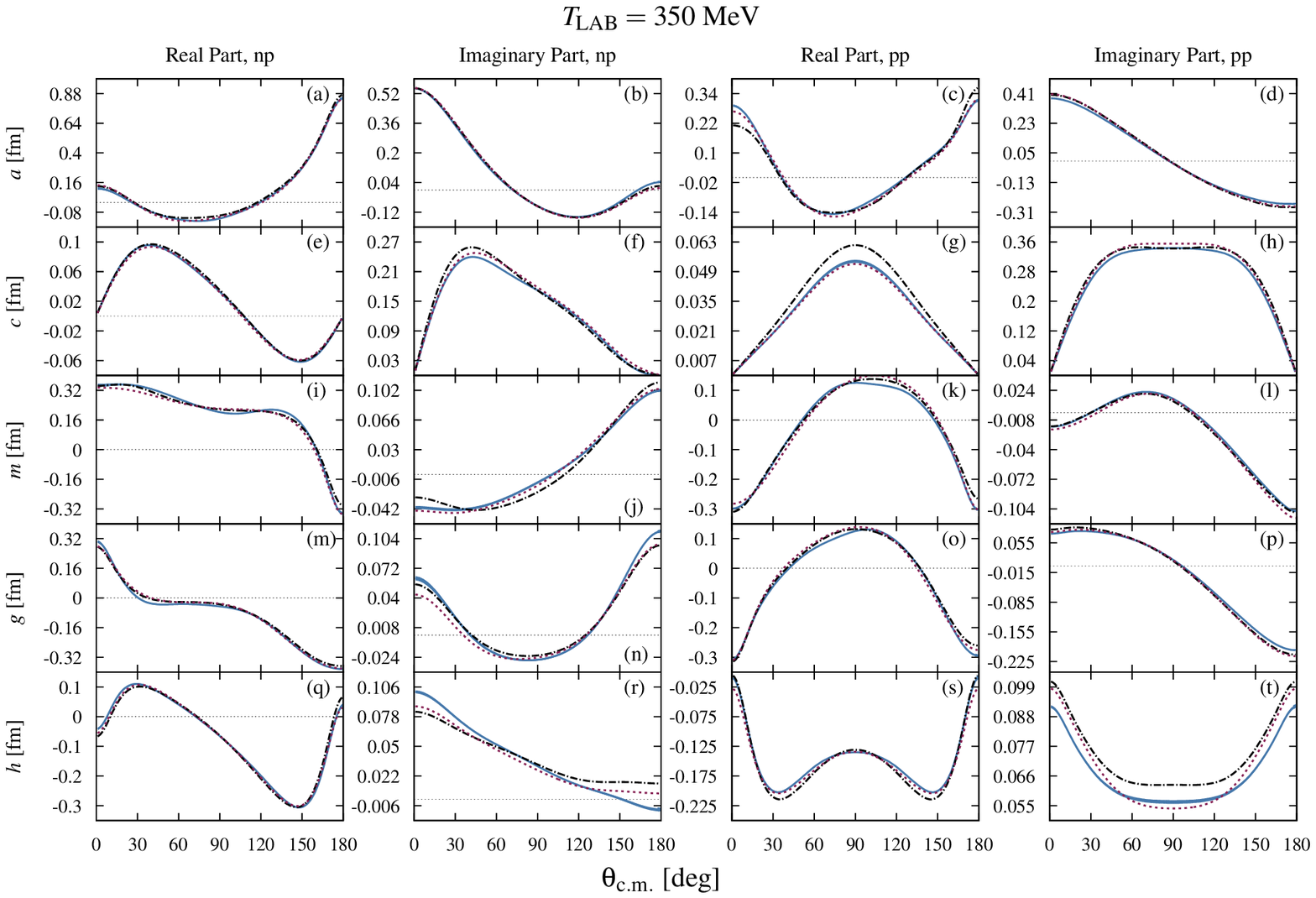,height=10cm,width=16cm}
\end{center}
\caption{Same as in Fig.~\ref{FigWolfenstein050} but for  $E_{\rm LAB}=350 {\rm
    MeV}$.}
\label{FigWolfenstein350}
\end{figure*}

\section{Numerical results}
\label{sec:3}

\subsection{Coarse graining EM interactions}

Of course, once we admit that the interaction below $r_c$ is unknown
there is no gain in directly extending the well-known charge-dependent
OPE tail for $r \le r_c$.  Unlike the purely strong piece of the NN
potential the electromagnetic contributions are known with much higher
accuracy and to shorter distances (see
e.g. Ref.~\cite{Wiringa:1994wb}) so that one might extend $V_{\rm
  em}(r)$ below $r_c$ adding a continuous contribution on top of the
delta-shells, so that the advantage of having a few radii in the
region $r \le r_c$ would be lost. To improve on this we coarse-grain
the EM interaction up to the pion production threshold. Thus, we look
for a discrete representation on the grid of the purely EM
contribution $V_{\rm em}(r)$, i.e. we take $\bar V_{\rm em} (r) =
\sum_n V_i^C \Delta r_i \delta (r-r_i) + \theta (r-r_c) V_{\rm em} (r) $,
where the $V_i^C$ are determined by reproducing the purely EM
scattering amplitude to high-precision and are not changed in the
fitting process. The result using the EM potential of
Ref.~\cite{Wiringa:1994wb} just turns out to involve the Coulomb
contribution in the central channel and the corresponding delta-shell
parameters $\lambda_i^C= V_i^C \Delta r_i M_p$ are given in the first line of
Table~\ref{tab:Fits}.  As expected from Nyquist sampling theorem, we
need at most $N=5$ sampling points which for simplicity are taken to
be equidistant with $\Delta r_i \equiv \Delta r = 0.6 {\rm fm}$
between the origin and $r_c=3 {\rm fm}$ to coarse grain the EM
interaction below $r \le r_c$. Thus we should have $V_i^{pp} =
V_i^{np} + V_i^C$ if charge symmetry was exact in strong interactions
for $r < r_c$, although some corrections are expected as documented
below.

\subsection{Fitting procedure}

In our fitting procedure we coarse grain the unknown short range part
of the interaction from the scattering data. We use the $
(\lambda_i)^{JS}_{l,l'} $'s as fitting parameters and minimize the
value of the $\chi^2$ using the Levenberg-Marquardt method where the
Hessian is computed explicitly~\cite{numrec-book}.  Actually, this is
a virtue of our delta-shell method which makes the computation of
derivatives with respect to the fitting parameters analytical and
straightforward.  As a consequence, explicit knowledge of the Hessian
allows to a faster search and finding of the minimum. 

We start with a complete
database compiling proton-proton and neutron-proton scattering data
obtained till 2007~\cite{NNonline,NNsaid,NNprovider}~\footnote{The
  most recent np fit to these data was carried out in
  Ref.~\cite{Gross:2008ps}.} and add two new data sets till
2013~\cite{Braun:2008eh,Daub:2012qb}. We carry out at any rate a
simultaneous pp and np fit for LAB kinetic energy below $350 {\rm
  MeV}$ to published data only. Unfortunately, some groups of these
data have a common but unknown normalization. We thus use the standard
floating~\cite{MacGregor:1968zzd} by including an additional
contribution to the $\chi^2$ as explained in detail, e.g., in
Ref.~\cite{Gross:2008ps}. The extra normalization data are labeled by
the subscript "norm'' below. We also apply the Nijmegen
PWA~\cite{Stoks:1993tb} $3\sigma$-criterion to reject possible
outliers from the main fit with a $3\sigma$-confidence level, a
strategy reducing the minimal $\chi^2$ but also enlarging the
uncertainties. Initially we consider $
N=2717|_{pp,exp}+151|_{pp,norm}+ 4734|_{np,exp}+262|_{np,norm}=
2868|_{pp}+4996|_{np} $ fitting data and get $\chi_{\rm min}^2=
3310|_{pp} + 8518|_{np} $ yielding $\chi^2/{\rm d.o.f.} =
1.51$. Applying the $3\sigma$-rejection and re-fitting the remaining
$N=2747|_{pp} + 3691|_{np}$ data we finally obtain $\chi^2_{\rm min}= 2813|_{pp}
+ 3985|_{np} $ yielding a total $\chi^2/{\rm d.o.f} = 1.06$.

While the linear relations of the $(\lambda_i)^{JS}_{l,l'}$ and
$V_{i,n}$ parameters are straightforward, limiting the number of
operators $O_n$ reduces the number of independent components of the
potential in the different partial waves. The fitting parameters
$(\lambda_n)^{JS}_{l,l'} $ entering the delta-shell potentials as
independent variables, Eq.~(\ref{eq:ds-pot}), are listed in
Table~\ref{tab:Fits} with their deduced uncertainties. All other
partial waves are consistently obtained from those using the linear relations 
between $(\lambda_i)^{JS}_{l,l'}$ and
$V_{i,n}$. Our final results
allow to fix the {\it same} pp and np potential parameters with the
exception of the central components of the potential as it is usually
the case in all joint pp+np analyses carried out so
far~\cite{Stoks:1993tb,Stoks:1994wp,Wiringa:1994wb,Machleidt:2000ge}.

We find that introducing more points or equivalently
reducing $\Delta r$ generates unnecessary correlations and does not
improve the fit. Also, lowering the value of $r_c$ below $3 {\rm fm}$,
requires overlapping the short-distance potential,
Eq.~(\ref{eq:ds-pot}), with the OPE plus EM corrections.
%Our potential is flexible enough as to display interesting features.
%For instance, 
We find that independent fits to pp and np, while
reducing each of the $\chi^2$-values, drive the minimum to
incompatible parameters and erroneous np phases in isovector
channels. Actually, the pp data constrain these channels most
efficiently and in a first step pp-fits where carried out to find
suitable starting parameters for the corresponding np-phases. Quite
generally, we have checked that the minimum is robust by proposing
several starting solutions. 

As a numerical check of our construction of the amplitudes we
reproduced the Wolfenstein parameters for the Reid93 and NijmII
potentials to high accuracy using $N=12000$ delta-shells grid points,
which ensures correctness of the strong contributions.  As a further
check of our implementation of the long-range EM effects along the
lines of Refs.~\cite{Stoks:1993tb,Durand:1957zz,Stoks:1990us} we have
also computed the $\chi^2/{\rm d.o.f.}$ for Reid93, NijmII and AV18 potentials
(fitted to data prior to 1993) which globally and bin-wise are
reasonably well reproduced when our database (coinciding with the one
of Ref.~\cite{Gross:2008ps} for np) includes only data prior to 1993.

\subsection{Comparing with other database}

In order to check the robustness of our database against other
selections of data we take the current SAID world
database~\cite{NNsaid} where unpublished data are also included and
some further data have been deleted from their analysis although the
total number exceeds our selected data.  If we consider these $N_{\rm
  SAID}=3061|_{pp, exp}+188|_{pp, norm} +4147|_{np,
  exp}+411|_{np,norm} = 3249|_{pp}+4558|_{np}$ data (without including
their deleted data) we get for our main fit (without re-fitting) the
value $\chi^2/{ N_{\rm SAID}} = 1.65$~\footnote{We do not include 14
  data of total pp cross section as our theoretical model includes all
  long range EM effects with no screening and, as is well known, the
  calculation diverges.}. Applying the $3\sigma$-rejection to this
database we get $\chi^2/{ N_{\rm SAID}} = 1.04$. If instead we fit our
model to this data base we initially get $\chi^2/{\rm d.o.f.}|_{\rm
    SAID} = 1.31$ which after $3\sigma$ selection of data becomes
  $\chi^2/{\rm d.o.f.}|_{\rm SAID} = 1.04$.

\subsection{Error propagation}

We determine the deuteron properties by solving the bound state
problem in the $^3S_1-^3D_1$ channel using the corresponding
parameters listed in Table~\ref{tab:Fits}. The predictions are
presented in table~\ref{tab:DeuteronP} where our quoted errors are
obtained from propagating those of Table~\ref{tab:Fits} by using the
full covariance matrix among fitting parameters. The comparison with
experimental values or high quality potentials where the deuteron
binding energy is used as an input is
satisfactory~\cite{Stoks:1993tb,Stoks:1994wp,Wiringa:1994wb,Machleidt:2000ge,Gross:2008ps}.

The outcoming and tiny theoretical error bands for the Deuteron form
factors (see e.g.~\cite{Gilman:2001yh}) are depicted in
Fig.~\ref{FigFormFactors} and are almost invisible at the scale of the
figure.  The rather small discrepancy between our theoretical results
and experimental form factor data is statistically significant and
might be resolved by the inclusion of Meson Exchange Currents.  In
Fig.~\ref{fig:all-phases} we show the active pp and np phases in the
fit with their propagated errors and compare them with the
PWA~\cite{Stoks:1993tb} and the AV18 potential~\cite{Wiringa:1994wb}
which provided a $\chi^2/ {\rm d.o.f} \lesssim 1 $. Note that the
$J=1$ phases show some discrepancy at higher energies, particularly in
the $\epsilon_1$ phase, where it is about the difference between the
PWA and the AV18 potential. Likewise, in
Figs.~\ref{FigWolfenstein050},
\ref{FigWolfenstein100},\ref{FigWolfenstein200} and
\ref{FigWolfenstein350} we also show a similar comparison for the pp
and np Wolfenstein parameters for several LAB energies.

Finally, as the previous
analyses~\cite{Stoks:1993tb,Stoks:1994wp,Wiringa:1994wb,Machleidt:2000ge,Gross:2008ps}
and the present paper show the form of the potential is not unique
providing a source of systematic errors. A first step along these
lines has been undertaken in Ref.~\cite{NavarroPerez:2012qf}.  Thus,
the uncertainties will generally be larger than those of purely statistical
nature estimated here.

\section{Conclusions}
\label{sec:4}
To summarize, we have determined a high-quality proton-proton and
neutron-proton interaction from a simultaneous fit to scattering data
and the deuteron binding energy with $\chi^2/{\rm d.o.f.}=1.06$. Our
short range potential consists of a few delta-shells for the lowest
partial waves. In addition, charge-dependent electromagnetic
interactions and one pion exchange are implemented.  We provide error
estimates on our fitting parameters.  Futher details will be presented
elsewhere.

\vskip.5cm
\acknowledgements 

We warmly thank Franz Gross for useful communications and
  providing data files. We also thank R. Schiavilla and R. Machleidt
  for communications. This work is partially supported by Spanish DGI
  (grant FIS2011-24149) and Junta de Andaluc{\'{\i}a} (grant FQM225).
  R.N.P. is supported by a Mexican CONACYT grant.

%\input{tables2}

%\input{tables3}

%\bibliography{amplitude_refs,refs}

\begin{thebibliography}{10}%
\makeatletter
\providecommand \@ifxundefined [1]{%
 \ifx #1\undefined \expandafter \@firstoftwo
 \else \expandafter \@secondoftwo
\fi
}%
\providecommand \@ifnum [1]{%
 \ifnum #1\expandafter \@firstoftwo
 \else \expandafter \@secondoftwo
\fi
}%
\providecommand \enquote [1]{``#1''}%
\providecommand \bibnamefont  [1]{#1}%
\providecommand \bibfnamefont [1]{#1}%
\providecommand \citenamefont [1]{#1}%
\providecommand\href[0]{\@sanitize\@href}%
\providecommand\@href[1]{\endgroup\@@startlink{#1}\endgroup\@@href}%
\providecommand\@@href[1]{#1\@@endlink}%
\providecommand \@sanitize [0]{\begingroup\catcode`\&12\catcode`\#12\relax}%
\@ifxundefined \pdfoutput {\@firstoftwo}{%
 \@ifnum{\z@=\pdfoutput}{\@firstoftwo}{\@secondoftwo}%
}{%
 \providecommand\@@startlink[1]{\leavevmode}%
 \providecommand\@@endlink[0]{}%
}{%
 \providecommand\@@startlink[1]{%
  \leavevmode
  \pdfstartlink
   attr{/Border[0 0 1 ]/H/I/C[0 1 1]}%
   user{/Subtype/Link/A<</Type/Action/S/URI/URI(#1)>>}%
  \relax
 }%
 \providecommand\@@endlink[0]{\pdfendlink}%
}%
\providecommand \url  [0]{\begingroup\@sanitize \@url }%
\providecommand \@url [1]{\endgroup\@href {#1}{\urlprefix}}%
\providecommand \urlprefix [0]{URL }%
\providecommand \Eprint[0]{\href }%
\@ifxundefined \urlstyle {%
  \providecommand \doi [1]{doi:\discretionary{}{}{}#1}%
}{%
  \providecommand \doi [0]{doi:\discretionary{}{}{}\begingroup
  \urlstyle{rm}\Url }%
}%
\providecommand \doibase [0]{http://dx.doi.org/}%
\providecommand \Doi[1]{\href{\doibase#1}}%
\providecommand \bibAnnote [3]{%
  \BibitemShut{#1}%
  \begin{quotation}\noindent
    \textsc{Key:}\ #2\\\textsc{Annotation:}\ #3%
  \end{quotation}%
}%
\providecommand \bibAnnoteFile [2]{%
  \IfFileExists{#2}{\bibAnnote {#1} {#2} {\input{#2}}}{}%
}%
\providecommand \typeout [0]{\immediate \write \m@ne }%
\providecommand \selectlanguage [0]{\@gobble}%
\providecommand \bibinfo [0]{\@secondoftwo}%
\providecommand \bibfield [0]{\@secondoftwo}%
\providecommand \translation [1]{[#1]}%
\providecommand \BibitemOpen[0]{}%
\providecommand \bibitemStop [0]{}%
\providecommand \bibitemNoStop [0]{.\EOS\space}%
\providecommand \EOS [0]{\spacefactor3000\relax}%
\providecommand \BibitemShut [1]{\csname bibitem#1\endcsname}%
%</preamble>
\bibitem{Machleidt:1989tm}%
  \BibitemOpen
  \bibfield{author}{%
  \bibinfo {author} {\bibfnamefont{R.}~\bibnamefont{Machleidt}},\ }%
  \bibfield{journal}{%
  \bibinfo {journal} {Adv.Nucl.Phys.}\ }%
  \textbf{\bibinfo {volume} {19}},\ \bibinfo {pages} {189} (\bibinfo {year}
  {1989})%
  \bibAnnoteFile{NoStop}{Machleidt:1989tm}%
%%CITATION = ANUPB,19,189;%%
\bibitem{Machleidt:2011zz}%
  \BibitemOpen
  \bibfield{author}{%
  \bibinfo {author} {\bibfnamefont{R.}~\bibnamefont{Machleidt}}\ and\ \bibinfo
  {author} {\bibfnamefont{D.}~\bibnamefont{Entem}},\ }%
  \bibfield{journal}{%
  \Doi{10.1016/j.physrep.2011.02.001}{\bibinfo {journal} {Phys.Rept.}}\ }%
  \textbf{\bibinfo {volume} {503}},\ \bibinfo {pages} {1} (\bibinfo {year}
  {2011})%
  \bibAnnoteFile{NoStop}{Machleidt:2011zz}%
\bibitem{glöckle1983quantum}%
  \BibitemOpen
  \bibfield{author}{%
  \bibinfo {author} {\bibfnamefont{W.}~\bibnamefont{Gl{\"o}ckle}},\ }%
  \emph{\bibinfo {title} {The quantum mechanical few-body problem}}\ (\bibinfo
  {publisher} {Springer-Verlag},\ \bibinfo {year} {1983})%
  \bibAnnoteFile{NoStop}{glöckle1983quantum}%
\bibitem{Arndt:2007qn}%
  \BibitemOpen
  \bibfield{author}{%
  \bibinfo {author} {\bibfnamefont{R.}~\bibnamefont{Arndt}}, \bibinfo {author}
  {\bibfnamefont{W.}~\bibnamefont{Briscoe}}, \bibinfo {author}
  {\bibfnamefont{I.}~\bibnamefont{Strakovsky}},\ and\ \bibinfo {author}
  {\bibfnamefont{R.}~\bibnamefont{Workman}},\ }%
  \bibfield{journal}{%
  \Doi{10.1103/PhysRevC.76.025209}{\bibinfo {journal} {Phys.Rev.}}\ }%
  \textbf{\bibinfo {volume} {C76}},\ \bibinfo {pages} {025209} (\bibinfo {year}
  {2007}),\ \Eprint{http://arxiv.org/abs/0706.2195}{arXiv:0706.2195 [nucl-th]}%
  \bibAnnoteFile{NoStop}{Arndt:2007qn}%
%%CITATION = ARXIV:0706.2195;%%
\bibitem{Stoks:1993tb}%
  \BibitemOpen
  \bibfield{author}{%
  \bibinfo {author} {\bibfnamefont{V.~G.~J.}\ \bibnamefont{Stoks}}, \bibinfo
  {author} {\bibfnamefont{R.~A.~M.}\ \bibnamefont{Kompl}}, \bibinfo {author}
  {\bibfnamefont{M.~C.~M.}\ \bibnamefont{Rentmeester}},\ and\ \bibinfo {author}
  {\bibfnamefont{J.~J.}\ \bibnamefont{de~Swart}},\ }%
  \bibfield{journal}{%
  \Doi{10.1103/PhysRevC.48.792}{\bibinfo {journal} {Phys. Rev.}}\ }%
  \textbf{\bibinfo {volume} {C48}},\ \bibinfo {pages} {792} (\bibinfo {year}
  {1993})%
  \bibAnnoteFile{NoStop}{Stoks:1993tb}%
%%CITATION = PHRVA,C48,792;%%
\bibitem{Stoks:1994wp}%
  \BibitemOpen
  \bibfield{author}{%
  \bibinfo {author} {\bibfnamefont{V.~G.~J.}\ \bibnamefont{Stoks}}, \bibinfo
  {author} {\bibfnamefont{R.~A.~M.}\ \bibnamefont{Klomp}}, \bibinfo {author}
  {\bibfnamefont{C.~P.~F.}\ \bibnamefont{Terheggen}},\ and\ \bibinfo {author}
  {\bibfnamefont{J.~J.}\ \bibnamefont{de~Swart}},\ }%
  \bibfield{journal}{%
  \Doi{10.1103/PhysRevC.49.2950}{\bibinfo {journal} {Phys. Rev.}}\ }%
  \textbf{\bibinfo {volume} {C49}},\ \bibinfo {pages} {2950} (\bibinfo {year}
  {1994}),\
  \Eprint{http://arxiv.org/abs/nucl-th/9406039}{arXiv:nucl-th/9406039}%
  \bibAnnoteFile{NoStop}{Stoks:1994wp}%
%%CITATION = NUCL-TH/9406039;%%
\bibitem{Wiringa:1994wb}%
  \BibitemOpen
  \bibfield{author}{%
  \bibinfo {author} {\bibfnamefont{R.~B.}\ \bibnamefont{Wiringa}}, \bibinfo
  {author} {\bibfnamefont{V.~G.~J.}\ \bibnamefont{Stoks}},\ and\ \bibinfo
  {author} {\bibfnamefont{R.}~\bibnamefont{Schiavilla}},\ }%
  \bibfield{journal}{%
  \Doi{10.1103/PhysRevC.51.38}{\bibinfo {journal} {Phys. Rev.}}\ }%
  \textbf{\bibinfo {volume} {C51}},\ \bibinfo {pages} {38} (\bibinfo {year}
  {1995}),\
  \Eprint{http://arxiv.org/abs/nucl-th/9408016}{arXiv:nucl-th/9408016}%
  \bibAnnoteFile{NoStop}{Wiringa:1994wb}%
%%CITATION = NUCL-TH/9408016;%%
\bibitem{Machleidt:2000ge}%
  \BibitemOpen
  \bibfield{author}{%
  \bibinfo {author} {\bibfnamefont{R.}~\bibnamefont{Machleidt}},\ }%
  \bibfield{journal}{%
  \Doi{10.1103/PhysRevC.63.024001}{\bibinfo {journal} {Phys. Rev.}}\ }%
  \textbf{\bibinfo {volume} {C63}},\ \bibinfo {pages} {024001} (\bibinfo {year}
  {2001}),\
  \Eprint{http://arxiv.org/abs/nucl-th/0006014}{arXiv:nucl-th/0006014}%
  \bibAnnoteFile{NoStop}{Machleidt:2000ge}%
%%CITATION = NUCL-TH/0006014;%%
\bibitem{Gross:2008ps}%
  \BibitemOpen
  \bibfield{author}{%
  \bibinfo {author} {\bibfnamefont{F.}~\bibnamefont{Gross}}\ and\ \bibinfo
  {author} {\bibfnamefont{A.}~\bibnamefont{Stadler}},\ }%
  \bibfield{journal}{%
  \Doi{10.1103/PhysRevC.78.014005}{\bibinfo {journal} {Phys.Rev.}}\ }%
  \textbf{\bibinfo {volume} {C78}},\ \bibinfo {pages} {014005} (\bibinfo {year}
  {2008})%
  \bibAnnoteFile{NoStop}{Gross:2008ps}%
\bibitem{taylor-book}%
  \BibitemOpen
  \bibfield{author}{%
  \bibinfo {author} {\bibfnamefont{J.~R.}\ \bibnamefont{Taylor}},\ }%
  \emph{\bibinfo {title} {An Introduction to Error Analysis: The Study of
  Uncertainties in Physical Measurements}}\ (\bibinfo {publisher} {University
  Science Books},\ \bibinfo {year} {1997})%
  \bibAnnoteFile{NoStop}{taylor-book}%
\bibitem{VanDerLeun1982261}%
  \BibitemOpen
  \bibfield{author}{%
  \bibinfo {author} {\bibfnamefont{C.~V.~D.}\ \bibnamefont{Leun}}\ and\
  \bibinfo {author} {\bibfnamefont{C.}~\bibnamefont{Alderliesten}},\ }%
  \bibfield{journal}{%
  \Doi{DOI: 10.1016/0375-9474(82)90105-1}{\bibinfo {journal} {Nucl. Phys.}}\ }%
  \textbf{\bibinfo {volume} {A380}},\ \bibinfo {pages} {261 } (\bibinfo {year}
  {1982})%
  \bibAnnoteFile{NoStop}{VanDerLeun1982261}%
\bibitem{Borbély198517}%
  \BibitemOpen
  \bibfield{author}{%
  \bibinfo {author} {\bibfnamefont{I.}~\bibnamefont{Borbély}}, \bibinfo
  {author} {\bibfnamefont{W.}~\bibnamefont{Grüebler}}, \bibinfo {author}
  {\bibfnamefont{V.}~\bibnamefont{König}}, \bibinfo {author}
  {\bibfnamefont{P.~A.}\ \bibnamefont{Schmelzbach}},\ and\ \bibinfo {author}
  {\bibfnamefont{A.~M.}\ \bibnamefont{Mukhamedzhanov}},\ }%
  \bibfield{journal}{%
  \Doi{DOI: 10.1016/0370-2693(85)91459-5}{\bibinfo {journal} {Phys. Lett.}}\ }%
  \textbf{\bibinfo {volume} {160B}},\ \bibinfo {pages} {17 } (\bibinfo {year}
  {1985})%
  \bibAnnoteFile{NoStop}{Borbély198517}%
\bibitem{Rodning:1990zz}%
  \BibitemOpen
  \bibfield{author}{%
  \bibinfo {author} {\bibfnamefont{N.~L.}\ \bibnamefont{Rodning}}\ and\
  \bibinfo {author} {\bibfnamefont{L.~D.}\ \bibnamefont{Knutson}},\ }%
  \bibfield{journal}{%
  \Doi{10.1103/PhysRevC.41.898}{\bibinfo {journal} {Phys. Rev.}}\ }%
  \textbf{\bibinfo {volume} {C41}},\ \bibinfo {pages} {898} (\bibinfo {year}
  {1990})%
  \bibAnnoteFile{NoStop}{Rodning:1990zz}%
%%CITATION = PHRVA,C41,898;%%
\bibitem{Klarsfeld1986373}%
  \BibitemOpen
  \bibfield{author}{%
  \bibinfo {author} {\bibfnamefont{S.}~\bibnamefont{Klarsfeld}}, \bibinfo
  {author} {\bibfnamefont{J.}~\bibnamefont{Martorell}}, \bibinfo {author}
  {\bibfnamefont{J.~A.}\ \bibnamefont{Oteo}}, \bibinfo {author}
  {\bibfnamefont{M.}~\bibnamefont{Nishimura}},\ and\ \bibinfo {author}
  {\bibfnamefont{D.~W.~L.}\ \bibnamefont{Sprung}},\ }%
  \bibfield{journal}{%
  \Doi{DOI: 10.1016/0375-9474(86)90400-8}{\bibinfo {journal} {Nucl. Phys.}}\ }%
  \textbf{\bibinfo {volume} {A456}},\ \bibinfo {pages} {373 } (\bibinfo {year}
  {1986})%
  \bibAnnoteFile{NoStop}{Klarsfeld1986373}%
\bibitem{Bishop:1979zz}%
  \BibitemOpen
  \bibfield{author}{%
  \bibinfo {author} {\bibfnamefont{D.~M.}\ \bibnamefont{Bishop}}\ and\ \bibinfo
  {author} {\bibfnamefont{L.~M.}\ \bibnamefont{Cheung}},\ }%
  \bibfield{journal}{%
  \Doi{10.1103/PhysRevA.20.381}{\bibinfo {journal} {Phys. Rev.}}\ }%
  \textbf{\bibinfo {volume} {A20}},\ \bibinfo {pages} {381} (\bibinfo {year}
  {1979})%
  \bibAnnoteFile{NoStop}{Bishop:1979zz}%
%%CITATION = PHRVA,A20,381;%%
\bibitem{deSwart:1995ui}%
  \BibitemOpen
  \bibfield{author}{%
  \bibinfo {author} {\bibfnamefont{J.~J.}\ \bibnamefont{de~Swart}}, \bibinfo
  {author} {\bibfnamefont{C.~P.~F.}\ \bibnamefont{Terheggen}},\ and\ \bibinfo
  {author} {\bibfnamefont{V.~G.~J.}\ \bibnamefont{Stoks}}}%
   (\bibinfo {year} {1995}),\
  \Eprint{http://arxiv.org/abs/nucl-th/9509032}{arXiv:nucl-th/9509032}%
  \bibAnnoteFile{NoStop}{deSwart:1995ui}%
%%CITATION = NUCL-TH/9509032;%%
\bibitem{Calogero:1965}%
  \BibitemOpen
  \bibfield{author}{%
  \bibinfo {author} {\bibfnamefont{F.}~\bibnamefont{Calogero}},\ }%
  \emph{\bibinfo {title} {Variable Phase Approach to Potential Scattering}}\
  (\bibinfo {publisher} {Academic Press, New York},\ \bibinfo {year} {1967})%
  \bibAnnoteFile{NoStop}{Calogero:1965}%
\bibitem{Aviles:1973ee}%
  \BibitemOpen
  \bibfield{author}{%
  \bibinfo {author} {\bibfnamefont{J.~B.}\ \bibnamefont{Aviles}},\ }%
  \bibfield{journal}{%
  \Doi{10.1103/PhysRevC.6.1467}{\bibinfo {journal} {Phys. Rev.}}\ }%
  \textbf{\bibinfo {volume} {C6}},\ \bibinfo {pages} {1467} (\bibinfo {year}
  {1972})%
  \bibAnnoteFile{NoStop}{Aviles:1973ee}%
%%CITATION = PHRVA,C6,1467;%%
\bibitem{PavonValderrama:2005ku}%
  \BibitemOpen
  \bibfield{author}{%
  \bibinfo {author} {\bibfnamefont{M.}~\bibnamefont{Pavon~Valderrama}}\ and\
  \bibinfo {author} {\bibfnamefont{E.~R.}\ \bibnamefont{Arriola}},\ }%
  \bibfield{journal}{%
  \Doi{10.1103/PhysRevC.72.044007}{\bibinfo {journal} {Phys. Rev.}}\ }%
  \textbf{\bibinfo {volume} {C72}},\ \bibinfo {pages} {044007} (\bibinfo {year}
  {2005})%
  \bibAnnoteFile{NoStop}{PavonValderrama:2005ku}%
%%CITATION = PHRVA,C72,044007;%%
\bibitem{Entem:2007jg}%
  \BibitemOpen
  \bibfield{author}{%
  \bibinfo {author} {\bibfnamefont{D.}~\bibnamefont{Entem}}, \bibinfo {author}
  {\bibfnamefont{E.}~\bibnamefont{Ruiz~Arriola}}, \bibinfo {author}
  {\bibfnamefont{M.}~\bibnamefont{Pavon~Valderrama}},\ and\ \bibinfo {author}
  {\bibfnamefont{R.}~\bibnamefont{Machleidt}},\ }%
  \bibfield{journal}{%
  \Doi{10.1103/PhysRevC.77.044006}{\bibinfo {journal} {Phys.Rev.}}\ }%
  \textbf{\bibinfo {volume} {C77}},\ \bibinfo {pages} {044006} (\bibinfo {year}
  {2008})%
  \bibAnnoteFile{NoStop}{Entem:2007jg}%
\bibitem{NavarroPerez:2011fm}%
  \BibitemOpen
  \bibfield{author}{%
  \bibinfo {author} {\bibfnamefont{R.}~\bibnamefont{Navarro~Perez}}, \bibinfo
  {author} {\bibfnamefont{J.}~\bibnamefont{Amaro}},\ and\ \bibinfo {author}
  {\bibfnamefont{E.}~\bibnamefont{Ruiz~Arriola}},\ }%
  \bibfield{journal}{%
  \Doi{10.1016/j.ppnp.2011.12.044}{\bibinfo {journal} {Prog.Part.Nucl.Phys.}}\
  }%
  \textbf{\bibinfo {volume} {67}},\ \bibinfo {pages} {359} (\bibinfo {year}
  {2012}),\ \Eprint{http://arxiv.org/abs/1111.4328}{arXiv:1111.4328 [nucl-th]}%
  \bibAnnoteFile{NoStop}{NavarroPerez:2011fm}%
%%CITATION = ARXIV:1111.4328;%%
\bibitem{Durand:1957zz}%
  \BibitemOpen
  \bibfield{author}{%
  \bibinfo {author} {\bibfnamefont{L.}~\bibnamefont{Durand}},\ }%
  \bibfield{journal}{%
  \Doi{10.1103/PhysRev.108.1597}{\bibinfo {journal} {Phys.Rev.}}\ }%
  \textbf{\bibinfo {volume} {108}},\ \bibinfo {pages} {1597} (\bibinfo {year}
  {1957})%
  \bibAnnoteFile{NoStop}{Durand:1957zz}%
%%CITATION = PHRVA,108,1597;%%
\bibitem{Stoks:1990us}%
  \BibitemOpen
  \bibfield{author}{%
  \bibinfo {author} {\bibfnamefont{V.}~\bibnamefont{Stoks}}\ and\ \bibinfo
  {author} {\bibfnamefont{J.}~\bibnamefont{De~Swart}},\ }%
  \bibfield{journal}{%
  \Doi{10.1103/PhysRevC.42.1235}{\bibinfo {journal} {Phys.Rev.}}\ }%
  \textbf{\bibinfo {volume} {C42}},\ \bibinfo {pages} {1235} (\bibinfo {year}
  {1990})%
  \bibAnnoteFile{NoStop}{Stoks:1990us}%
%%CITATION = PHRVA,C42,1235;%%
\bibitem{numrec-book}%
  \BibitemOpen
  \bibfield{author}{%
  \bibinfo {author} {\bibfnamefont{W.~H.}\ \bibnamefont{Press}},\ % 
  \bibinfo {author} {\bibfnamefont{S.~A.}\ \bibnamefont{Teukolsky}},\ %
  \bibinfo {author} {\bibfnamefont{W.~T.}\ \bibnamefont{Vetterling}},\ %
  \bibinfo {author} {\bibfnamefont{B.~P.}\ \bibnamefont{Flannery}},\ }%
  \emph{\bibinfo {title} { Numerical recipes 3rd edition: The art of scientific computing}}\ (\bibinfo {publisher} {Cambridge university press},\ \bibinfo {year} {2007})%
  \bibAnnoteFile{NoStop}{numrec-book}%
\bibitem{NNonline}%
  \BibitemOpen
  \ \url{http://nn-online.org/}%
  \bibAnnoteFile{NoStop}{NNonline}%
\bibitem{NNsaid}%
  \BibitemOpen
  \ \url{http://gwdac.phys.gwu.edu/analysis/nn_analysis.html}%
  \bibAnnoteFile{NoStop}{NNsaid}%
\bibitem{NNprovider}%
  \BibitemOpen
  \
  \href{https://play.google.com/store/apps/details?id=es.ugr.amaro.nnprovider}
{NNprovider}
https://play.google.com/store/apps/details?id=
es.ugr.amaro.nnprovider
%\bibAnnoteFile{NoStop}{NNprovider}%

%\cite{Braun:2008eh}
\bibitem{Braun:2008eh} 
  R.~T.~Braun, W.~Tornow, C.~R.~Howell, D.~E.~Gonzalez Trotter, C.~D.~Roper, F.~Salinas, H.~R.~Setze and R.~L.~Walter {\it et al.},
  %``Neutron-proton analyzing power at 12-MeV and inconsistencies in parametrizations of nucleon-nucleon data,''
  Phys.\ Lett.\ B {\bf 660}, 161 (2008)
  [arXiv:0801.4600 [nucl-ex]].
  %%CITATION = ARXIV:0801.4600;%%
  %3 citations counted in INSPIRE as of 28 May 2013


%\cite{Daub:2012qb}
\bibitem{Daub:2012qb} 
  B.~H.~Daub, V.~Henzl, M.~A.~Kovash, J.~L.~Matthews, Z.~W.~Miller, K.~Shoniyozov and H.~Yang,
  %``Measurement of the Neutron-Proton and Neutron-Carbon Total Cross Section from 150 to 800 keV,''
  Phys.\ Rev.\ C {\bf 87}, 014005 (2013)
  [arXiv:1211.2203 [nucl-ex]].
  %%CITATION = ARXIV:1211.2203;%%


%\cite{MacGregor:1968zzd}
\bibitem{MacGregor:1968zzd}
  M.~H.~MacGregor, R.~A.~Arndt and R.~M.~Wright,
  %``Determination of the Nucleon-Nucleon Scattering Matrix. VII. (p, p) Analysis from 0 to 400 MeV,''
  Phys.\ Rev.\  {\bf 169} (1968) 1128.
  %%CITATION = PHRVA,169,1128;%%
  %69 citations counted in INSPIRE as of 28 May 2013

\bibitem{Gilman:2001yh}%
  \BibitemOpen
  \bibfield{author}{%
  \bibinfo {author} {\bibfnamefont{R.~A.}\ \bibnamefont{Gilman}}\ and\ \bibinfo
  {author} {\bibfnamefont{F.}~\bibnamefont{Gross}},\ }%
  \bibfield{journal}{%
  \Doi{10.1088/0954-3899/28/4/201}{\bibinfo {journal} {J.Phys.}}\ }%
  \textbf{\bibinfo {volume} {G28}},\ \bibinfo {pages} {R37} (\bibinfo {year}
  {2002}),\ \Eprint{http://arxiv.org/abs/nucl-th/0111015}{arXiv:nucl-th/0111015
  [nucl-th]}%
  \bibAnnoteFile{NoStop}{Gilman:2001yh}%
%%CITATION = NUCL-TH/0111015;%%
\bibitem{NavarroPerez:2012qf}%
  \BibitemOpen
  \bibfield{author}{%
  \bibinfo {author} {\bibfnamefont{R.}~\bibnamefont{Navarro~Perez}}, \bibinfo
  {author} {\bibfnamefont{J.}~\bibnamefont{Amaro}},\ and\ \bibinfo {author}
  {\bibfnamefont{E.}~\bibnamefont{Ruiz~Arriola}}}
   (\bibinfo {year} {2012}),\
  \Eprint{http://arxiv.org/abs/1202.2689}{arXiv:1202.2689 [nucl-th]}, Phys.Lett. {\bf B} (to appear)%
  \bibAnnoteFile{NoStop}{NavarroPerez:2012qf}%
%%CITATION = ARXIV:1202.2689;%%
\end{thebibliography}

%Merlin.mbs v4.21 2009-07-09.
%

\end{document}